\documentclass[9pt,twocolumn,twoside]{osajnl}
\usepackage[justification=centering]{caption}
\usepackage{float}
\usepackage{flushend}

\journal{josab} 

\setboolean{shortarticle}{false}

\title{Phase-sensitive nonclassical properties in quantum metrology with
displaced squeezed vacuum state}

\author[1,2]{Zhiwei Tao}
\author[2,*]{Yichong Ren}
\author[2]{Azezigul Abdukirim}
\author[1,2]{Shiwei Liu}
\author[2]{Ruizhong Rao}

\affil[1]{School of Environmental Science and Optoelectronic Technology, University of
Science and Technology of China, Hefei 230022, China}
\affil[2]{Key Laboratory of Atmospheric Optics, Anhui Institute of Optics and Fine
Mechanics, Chinese Academy of Sciences, Hefei 230031, China}

\affil[*]{Corresponding author: rych@aiofm.ac.cn}

\doi{\url{http://dx.doi.org/10.1364/JOSAB.XX.XXXXXX}}

\begin{abstract}
We predict that the phase-dependent error distribution of locally unentangled quantum
states directly affects quantum parameter estimation accuracy. Therefore, we
employ the displaced squeezed vacuum (DSV) state as a probe state and
investigate an interesting question of the phase-sensitive nonclassical
properties in DSV's metrology. We found that the accuracy limit of parameter
estimation is a function of the phase-sensitive parameter $\phi -\theta /2$
with a period $\pi $. We show that when $\phi -\theta /2$ $\in \left[ k\pi
/2,3k\pi /4\right) \left( k\in 
\mathbb{Z}
\right) $, we can obtain the accuracy of parameter estimation approaching
the ultimate quantum limit through using the DSV state with the larger
displacement and squeezing strength, whereas $\phi -\theta /2$ $\in \left(
3k\pi /4,k\pi \right] \left( k\in 
\mathbb{Z}
\right) $, the optimal estimation accuracy can be acquired only when the DSV state
degenerates to squeezed-vacuum state.
\end{abstract}

\setboolean{displaycopyright}{true}

\begin{document}

\maketitle

\section{INTRODUCTION}

Quantum metrology, also known as quantum parameter estimation, primarily
focuses on enhancing the measurement accuracy of parameter estimation in the
dynamical evolution of the quantum probe system by using the features of
quantum mechanics, which demonstrates an impressive significance both in
theoretical predictions and experiments\cite{c1,c2,c3,c4}. At the early
stage, the improvement sought is an enhanced sensitivity achieved by a given
number of resources, such as number of probes, mean and maximum energy,
number of measurements and choice of integration time. However, without any
quantum instruments, the actual accuracy limit that can be achieved is
determined by the quantum noise of a single photon after quantization of the
electromagnetic field, in quantum optics typically so-called the shot-noise
limit\cite{c5}.

On the other hand, the optimal estimation accuracy is also bound by the
uncertainty principle of quantum mechanics, which easily translates into $1/N
$ scaling of Heisenberg limit\cite{c5} on parameter estimation using an $N$%
-photon state. To beat the shot-noise limit and constantly approach the
Heisenberg scaling, most commonly alternative quantum features such as
entanglement\cite{c4,c6,c7,c8,c9,c10,c11}, correlations\cite{c12,c13},
non-trivial Hamiltonians\cite{c14,c15,c16} and identical particles\cite%
{c17,c18} have been employed to improve measurement accuracy over past
decades substantially. Especially, many researchers have focused their
efforts primarily on how to find the optimal probe state and measurement for
estimating any parameter or multiple parameters encoded in a bosonic
Gaussian channel\cite{c5,c19,c20,c21,c22,c23,c24,c25,c26,c27,c28} and
achieve exactly the same estimation precision as entangled probes.

However, although they have elaborated general methods for improving the
parameter estimation accuracy extensively, very little research has been
conducted on another important physical issue: Where the parameter
estimation error arises? Notably, it is well-known that the role of the
unitary parameter operator makes the input state obtain a fixed parameter
shift. Therefore, we now have to pose the theoretically important question:
why does the estimated parameter have a standard deviation? We intuitively
believe that for unentangled local quantum states, the error that comes from
the estimated parameters might be partially determined by the initial phase
distribution of the input quantum state. In other words, the error contour's
shape of the input quantum state\cite{c29} might directly affect the
parameter estimation accuracy. Additionally, since quantum Fisher
information (QFI) is associated with the distinguishability of two
infinitely close-by quantum states\cite{c21}, we anticipate that both
combinations might jointly impact the accuracy limit of parameter estimation.

To examine our anticipation, in this work, we employ the displaced squeezed
vacuum (DSV) state as a probe state and investigate their phase-sensitive
nonclassical properties in quantum metrology. DSV state, as a special form
of single-mode Gaussian state, have several useful characteristics for our
exploration such as it contains the phase-dependent noise, reduced below
that of the coherent state for some phases and enhanced above that of the
coherent state for others\cite{c30}. Because of this phase-dependent error
distribution DSV state leads many technological applications, particularly
in the detection of weak signals\cite{c31,c32,c33}. Besides, in analogy to
quadrature squeezing, another form of squeezing such as number squeezing can
also be accomplished by rotating the DSV's phase $\phi -\theta /2$ coming
from the squeezing parameter $\xi =re^{i\theta }$ and displacement parameter 
$\alpha =\left\vert \alpha \right\vert e^{i\phi }$\cite{c30}, as depicted in
Fig. \ref{fig:1}.

Based on the above nonclassical properties, we first evaluate whether the
compound phase-sensitive parameter $\phi -\theta /2$ is connected to the
Cramer-Rao limit of the DSV state and then give a quantitative relationship
between $\phi -\theta /2$ and the QFI of the single-mode DSV state. Our
conclusion is that the accuracy limit of parameter estimation is a periodic
function of $\phi -\theta /2$ with a period $\pi $, which is influenced by,
but not uniquely influencing, the compound physical quantity. When $\phi
-\theta /2\in \left[ k\pi /2,3k\pi /4\right) \left( k\in 
\mathbb{Z}
\right) $, the DSV state with the larger squeezing strength and displacement
can achieve the most accurate parameter estimation performance, whereas in
the $\phi -\theta /2\in \left( 3k\pi /4,k\pi \right] \left( k\in 
\mathbb{Z}
\right) $ regime, the same effect can be obtained only when the DSV state
degenerates to a squeezed-vacuum state.

The paper is structured as follows. In Sec. \ref{sec:2}, we briefly introduce the
concept of QFI and define the phase-sensitive parameter $\phi -\theta /2$,
notably, $\phi -\theta /2$ keep invariance under the parameterization
process. In Sec. \ref{sec:3}, to rule out the overestimation of parameter information%
\cite{c34,c35,c36,c37}, we first calculate the QFI of the single-mode DSV
state after introducing an external parameter reference. Secondly, based on
the results of the DSV's QFI, we investigate how the accuracy limit of
parameter estimation changes with respect to the phase-sensitive parameter
under some typical displacement and squeezing parameters. Then, we analyze a
more general result about how these intrinsic DSV's parameters interplay the
accuracy limit of parameter estimation under different phase-sensitive
conditions and find what pair of DSV's intrinsic parameters gives the
optimal estimation accuracy under some typical phase-sensitive conditions. Our main
results are summarized in Sec. \ref{sec:4}.

\section{THEORY}
\label{sec:2}

\subsection{Quantum Fisher information}

Generally speaking, the procedure of parameter estimation can be roughly
decomposed into four steps: probe state preparation, parameterization or
dynamical evolution, measurement of the parameter encoded state as well as
estimation of parameter through postprocessing the measurement data. Thus,
from the above four steps, it is clearly shown to us that there exists an
optimal measurement strategy allowing the parameter estimation with the
highest accuracy. Assuming that the parameter to be estimated is one of the
variables $\varphi $ in the system, then the estimated limit is followed by
quantum Cramer-Rao inequality\cite{c38,c39}
\begin{equation}
\Delta ^{2}\varphi \geq \frac{1}{MI_{\varphi }}  \label{eq1}
\end{equation}%
\begin{figure}[htbp]
\centering
\includegraphics[width=\linewidth]{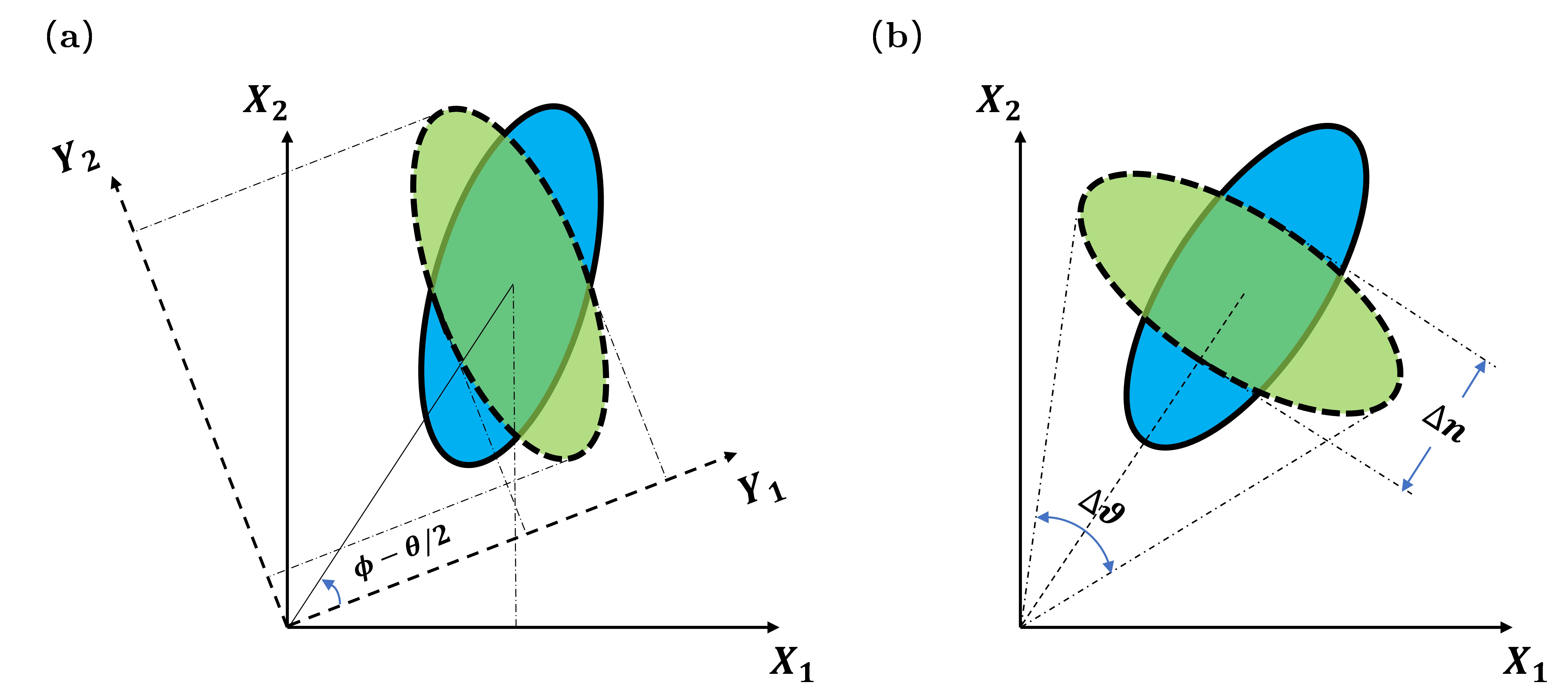}
\caption{Error ellipse in complex amplitude plane for the DSV state $D\left( \alpha
\right) S\left( \xi \right) \left\vert 0\right\rangle$, where $\xi =re^{i\theta }$ and $\alpha =\left\vert \alpha \right\vert
e^{i\phi }$. (a) Definition of phase-sensitive parameter $\phi -\theta /2$; (b) Rotated error ellipse with $\phi -\theta /2=k\pi /2\left( k\in \mathbb{Z} \right)$ and $\phi -\theta /2=k\pi \left( k\in \mathbb{Z} \right)$, where $\Delta \vartheta $ and $\Delta n$ represent the phase uncertainty and the amplitude uncertainty for given rotation angle respectively.}
\label{fig:1}
\end{figure}where $M$ is the number of measurements in experiments, $I_{\varphi }$
stands for QFI associated with the measurement accuracy and $\Delta
^{2}\varphi $ represents the variance of the parameter $\varphi $.

QFI is a natural generalization of the variance of the classical score
function, which measures the amount of information contained in a
parameterized random variable, defined as follows

\begin{equation}
I_{\varphi }=Tr\left( \rho _{\varphi }L_{\rho _{\varphi }}^{2}\right) 
\label{eq2}
\end{equation}%
where $\rho _{\varphi }$ is the probe state after parametrical encoding, $%
L_{\rho _{\varphi }}$ is the symmetric logarithmic derivative determined by $%
\partial _{\varphi }\rho _{\varphi }=1/2\left( L_{\rho _{\varphi }}\rho
_{\varphi }+\rho _{\varphi }L_{\rho _{\varphi }}\right) $. Another physical
meaning of QFI is that it geometrically measures the distinguishability
between two quantum states $\rho \left( \varphi \right) $ and $\rho \left(
\varphi +d\varphi \right) $ that differ infinitesimally in $d\varphi $,
which can also be expressed by the Bures distance\cite{c40}

\begin{equation}
I_{\varphi }=4ds_{Bures}^{2}\left( \rho \left( \varphi \right) ,\rho \left(
\varphi +d\varphi \right) \right) /d\varphi ^{2}  \label{eq3}
\end{equation}%
where $ds_{Bures}^{2}\left( \rho ,\sigma \right) \equiv 2\left[ 1-\sqrt{%
F\left( \rho ,\sigma \right) }\right] $ denotes the Bures distance and $%
F\left( \rho ,\sigma \right) \equiv \left[ Tr\left( \sqrt{\rho }\sigma \sqrt{%
\rho }\right) ^{1/2}\right] ^{2}$ is the Uhlmann fidelity.

For pure states $\rho _{\varphi }=\left\vert \psi _{\varphi }\right\rangle
\left\langle \psi _{\varphi }\right\vert $, the QFI simplifies to the
overlap of the derivative of the state with itself and the original state,
which can be described as\cite{c5}

\begin{equation}
I_{\varphi }=4\left( \langle \partial _{\varphi }\psi _{\varphi }\left\vert
\partial _{\varphi }\psi _{\varphi }\right\rangle +\left\vert \langle
\partial _{\varphi }\psi _{\varphi }\left\vert \psi _{\varphi }\right\rangle
\right\vert ^{2}\right)   \label{eq4}
\end{equation}%
where $\left\vert \partial _{\varphi }\psi _{\varphi }\right\rangle
=\partial \left\vert \psi _{\varphi }\right\rangle /\partial \varphi $.
Generally, parameterization for a given scenario is not always
straightforward and can lead to different results\cite{c34}. Therefore, to
prevent the exaggerated estimation of QFI, an external parameter reference
is invoked to allow a well-defined parameter $\varphi $. In the absence of
parameter reference, one has to pay attention to what parameter encoding
strategies are required to the corresponding experimental setup for
attaining the accurate QFI\cite{c41,c42,c43}.

Therefore, if we consider the input state $\left\vert \psi \right\rangle $
of two modes, the above procedure becomes a two-parameter estimation problem
and a QFI matrix needs to be employed. Assuming the two parameters
introduced is $\varphi _{1}$ and $\varphi _{2}$, under the basis $\varphi
_{\pm }=\varphi _{1}\pm \varphi _{2}$, the process of parametrical encoding
can be realized by\cite{c36} 
\begin{equation}
\left\vert \psi _{\varphi }\right\rangle =\exp \left[ i\left( \varphi
_{+}G_{+}+\varphi _{-}G_{-}\right) \right] \left\vert \psi \right\rangle 
\label{eq5}
\end{equation}%
where $G_{+}=\frac{1}{2}\left( \widehat{a}^{\dag }\widehat{a}+\widehat{b}%
^{\dag }\widehat{b}\right) $ and $G_{-}=\frac{1}{2}\left( \widehat{a}^{\dag }%
\widehat{a}-\widehat{b}^{\dag }\widehat{b}\right) $. $\widehat{a}$, $%
\widehat{b}$ denotes photon annihilation operators for the signal and the
idler. Since we are interested in the variance of $\varphi _{-}$, the
element of QFI matrix $I_{--}$ can be calculated following Eq. (\ref{eq4}) by%
\cite{c44}

\begin{equation}
I_{--}=4\mathbf{Var}\left( G_{-}\right) =4\left( \left\langle \psi _{\varphi
}\right\vert G_{-}^{2}\left\vert \psi _{\varphi }\right\rangle -\left\vert
\left\langle \psi _{\varphi }\right\vert G_{-}\left\vert \psi _{\varphi
}\right\rangle \right\vert ^{2}\right)  \label{eq6}
\end{equation}

\begin{figure}[htbp]
\centering
\includegraphics[width=0.75\linewidth]{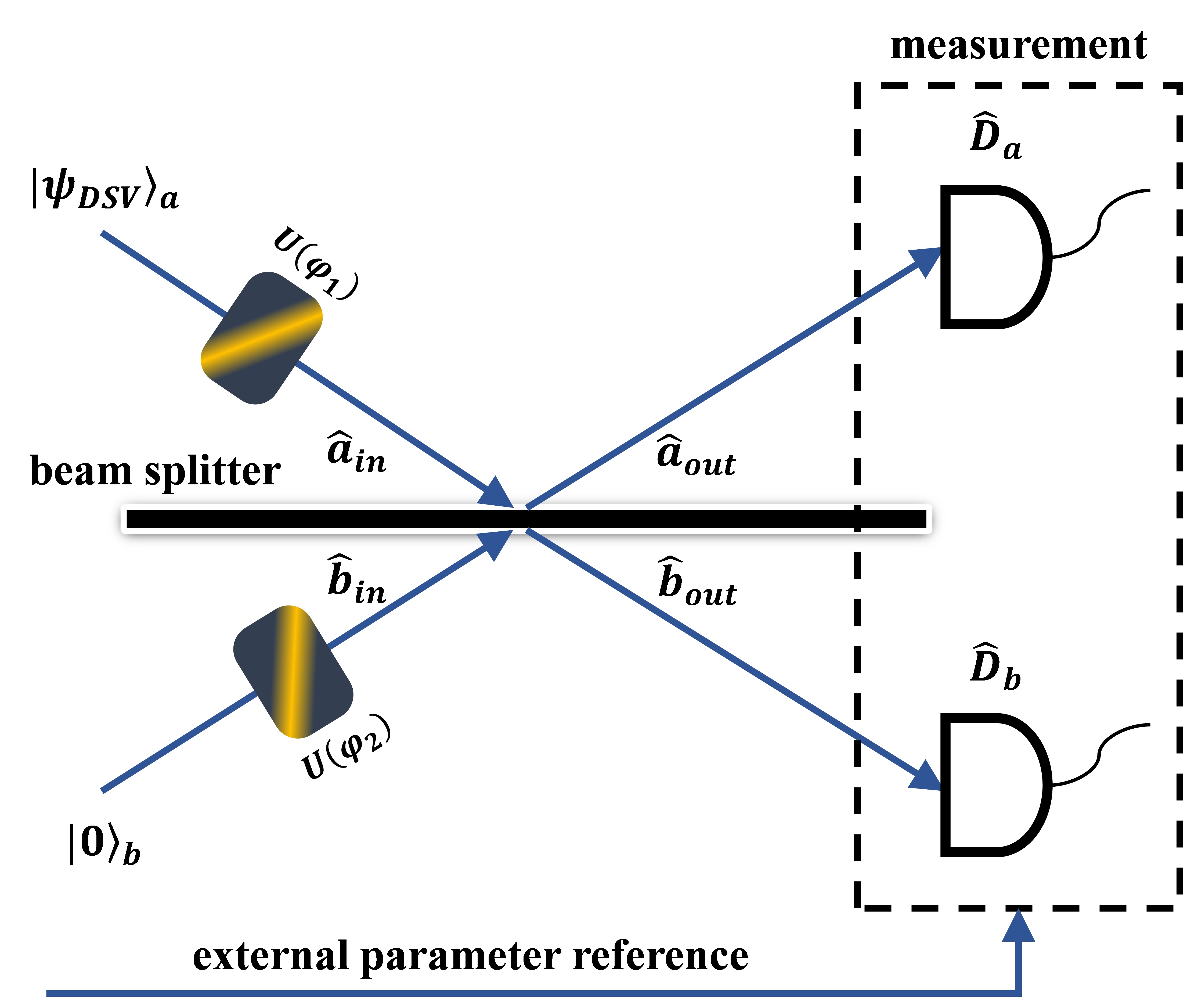}
\caption{Interferometric parameter estimation scheme
	for the DSV state. The difference between $\varphi_{1}-\varphi_{2}$ is estimated, based on the measurement strategies $\widehat{D}_{a}$ and $\widehat{D}_{b}$. The external parameter estimation is employed to prevent the overestimation of the DSV's QFI.}
\label{fig:2}
\end{figure}

\subsection{Phase-sensitive parameter of the DSV state}

A coherent state has half of unity identical uncertainty both in the
horizontal and vertical direction, which is the closest quantum state to the
classical one. Hence, to reduce the noise in one of two directions, the
common strategy can be realized by squeezing the wave pocket at the expense
of the corresponding increased fluctuations in the conjugate direction.
Theoretically, the squeezed state can be usually generated by the secondary
action of the squeeze operator $S\left( \xi \right) $ and the displacement
operator $D\left( \alpha \right) $ on the vacuum state%
\begin{equation}
\left\vert \psi \right\rangle \equiv S\left( \xi \right) D\left( \alpha
\right) \left\vert 0\right\rangle   \label{eq7}
\end{equation}%
where $\xi =re^{i\theta }$ and $\alpha =\left\vert \alpha \right\vert
e^{i\phi }$. $\theta $ and $\phi $ constitute the phase-sensitive parameter
shown in Fig. \ref{fig:1}(a). Another definition called the DSV state is considered in
this paper through exchanging the two unitary transformations imprinted on
the vacuum state. Notably, these two operators do not commute with each
other, which connect by the following relationship\cite{c30,c45}%
\begin{eqnarray}
S\left( \xi \right) D\left( \alpha \right)  &=&D\left( \beta \right) S\left(
\xi \right)   \label{eq8} \\
\beta  &=&\alpha \cosh r+\alpha ^{\ast }e^{i\theta }\sinh r  \notag
\end{eqnarray}

As shown in Fig. \ref{fig:1}(b), we observe the rotation of the DSV's error contour
leads to different uncertainties both in the phase and amplitude of electric
field (e.g., the green ellipse with dashed line has a maximum phase
uncertainty $\Delta \vartheta $ and minimum amplitude uncertainty $\Delta n$%
, nevertheless, the situation is reversed for the error contour
counterclockwise rotating $\pi /2$, see the blue one with full line). With
this in mind, we somewhat intuitively considered that selecting an
appropriate squeeze direction such as a decreased phase uncertainty may
facilitate us to obtain more accurate information about the unknown
parameters. Concretely, when the compound physical quantity $\phi -\theta /2$
(illustrate in Fig. \ref{fig:1}(a)), we called a phase-sensitive parameter, equals to
an integer multiple of $\pi /2$, the DSV state may achieve better
performance for parameter estimation compared with coherent states.
Conversely, the reversed conclusions may be acquired when $\phi -\theta /2=k\pi
\left( k\in 
\mathbb{Z}
\right) $.

\begin{figure*}[h]
\centering
\includegraphics[width=0.75\linewidth]{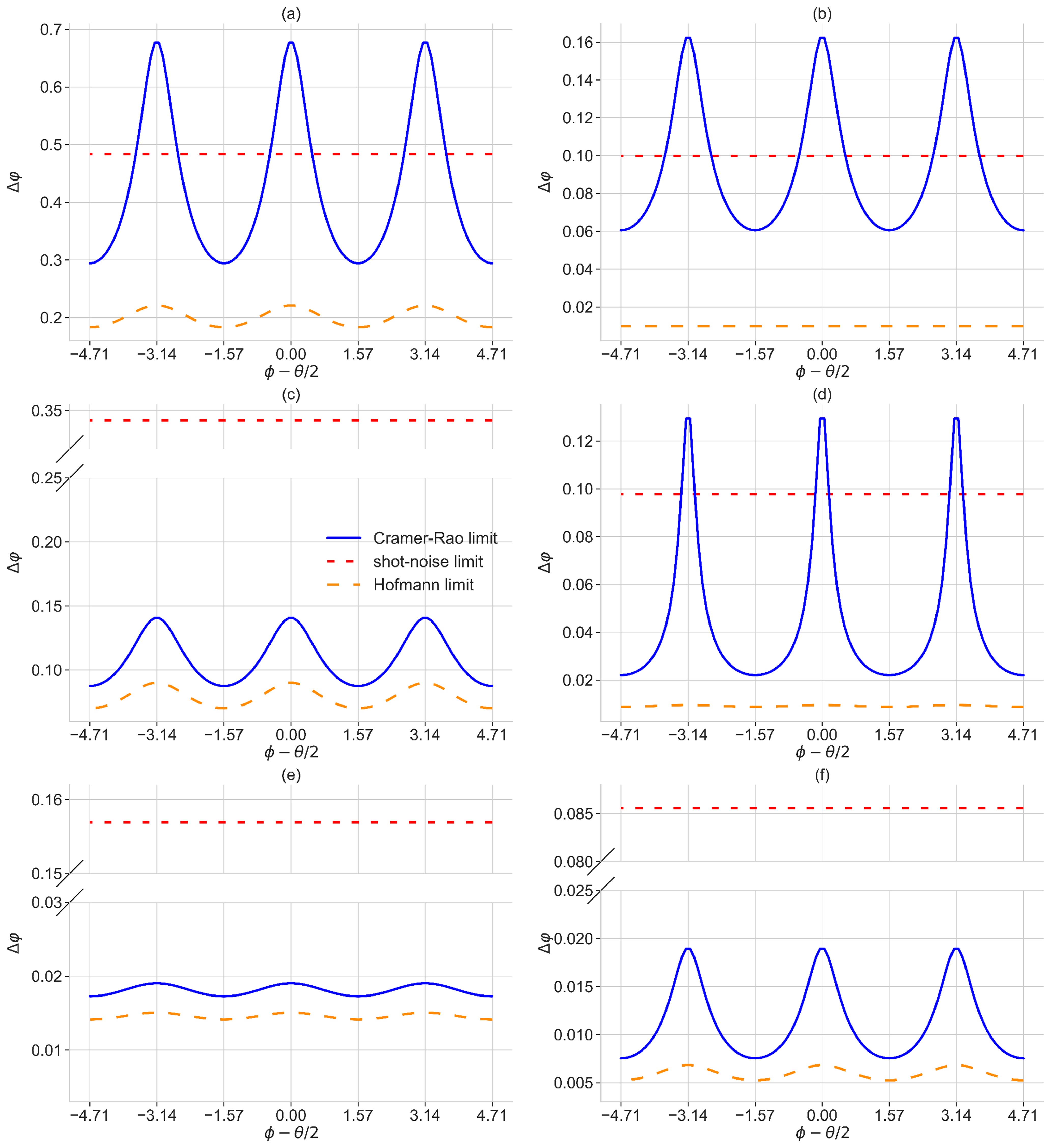}
\caption{Accuracy limit of parameter estimation as a function of the compound phase-sensitive parameter $\phi -\theta /2$ in the range $\left [-3\pi /2, 3\pi /2\right ]$ for different displacements and squeezing strengths (a) $r=0.5$, $\left\vert \alpha\right\vert =2$\ (b) $r=0.5$, $\left\vert \alpha\right\vert =10$\ (c) $r=1.5$, $\left\vert \alpha\right\vert =2$\ (d) $r=1.5$, $\left\vert \alpha\right\vert =10$\ (e)$r=2.5$, $\left\vert \alpha\right\vert =2$\ (f) $r=2.5$, $\left\vert \alpha\right\vert =10$.}
\label{fig:3}
\end{figure*}

\section{RESULTS}
\label{sec:3}
A standard Mach-Zehnder interferometer with two arms $a$ and $b$ usually
consists of two $50:50$ beam splitters and a phase shifter. The input
photons passing through the first beam splitter introduce a parameter to be
estimated. Then, these two photons interfere at the second beam splitter.
Finally, the magnitude of the parameter can be estimated from measurement
outcomes through data postprocessing. An important reason why such a scheme
effectively improves the accuracy of parameter estimation is that the first
beam splitter converts the incident photonic state into a path-entangled
state, thus enabling quantum-enhanced measurements.

In the present contribution, we consider an interferometer without including
the first beam splitter for achieving the same performance compared with
entanglement, the initial DSV state and the vacuum state prepared in mode $a$
and $b$, as shown in Fig. \ref{fig:2}, and acquired two phase shifts $\varphi _{1}$
and $\varphi _{2}$ in the channel $a$ and $b$ respectively. Undergoing the
parametrical encoding, the QFI of the output state $\left\vert \psi
_{\varphi }\right\rangle $ in Eq. (\ref{eq6}) can be expressed by

\begin{align}
I_{--}& =\sinh ^{2}r\cosh ^{2}r-\left\vert \alpha \right\vert ^{2}\left\{
2\sinh r\cosh r\cos \left[ 2\left( \phi -\theta /2\right) \right] -1\right\}
\label{eq9} \\
& +\left( 2\left\vert \alpha \right\vert ^{2}+1\right) \sinh ^{2}r+\sinh
^{4}r  \notag
\end{align}%
Substituting the above equation into Eq. (\ref{eq1}), we can achieve the
Cramer-Rao limit for the DSV state. It can be seen from the above equation
that the maximum accuracy of parameter estimation is a function of $\phi
-\theta /2$ with a period of $\pi $. when $\phi -\theta /2=k\pi \left( k\in 
\mathbb{Z}
\right) $, since $I_{--}$ obtains the minimum values, the worst accuracy
will be acquired at this moment. On the contrary, the reversed conclusions are
reported under a better phase-sensitive circumstance, which is the same as
our previous conjecture.

Before the formal discussion, we state that the above-mentioned Heisenberg
limit is commonly defined as the inverse of the average photon number $%
\overline{n}$ of the input state (i.e., $\Delta \varphi_{HL} =1/\overline{n}$).
However, it should be noted that the ultimate accuracy of parameter
estimation will be underestimated if the photon number fluctuations are
neglected, especially in the high fluctuations regime. Therefore, to prevent
this underestimation, a more direct definition of the ultimate quantum limit
in Hofmann\cite{c46} typically for scaling as $1/\sqrt{\overline{n^{2}}}$
with averaged squared photon numbers $\overline{n^{2}}$

\begin{eqnarray}
\overline{n^{2}} &=&\left\vert \alpha \right\vert ^{4}-\left\vert \alpha
\right\vert ^{2}\left\{ 2\sinh r\cosh r\cos \left[ 2\left( \phi -\theta
/2\right) \right] -1\right\}  \notag \\
&&+\sinh ^{2}r\cosh ^{2}r+\left( 4\left\vert \alpha \right\vert
^{2}+1\right) \sinh ^{2}r  \label{eq10} \\
&&+2\sinh ^{4}r  \notag
\end{eqnarray}

\subsection{Under the same phase-sensitive condition}

The comparison of the Cramer-Rao, the Hofmann, and the shot-noise limit as a
function of $\phi -\theta /2$ is presented in Fig. \ref{fig:3} for different typical
squeezing strengths and displacements. As shown in the top row of Fig. \ref{fig:3}, we
observe that in the $\phi -\theta /2=k\pi /2\left( k\in 
\mathbb{Z}
\right) $ regime, the Cramer-Rao limit of the DSV state beats the $1/\sqrt{%
\overline{n}}$ scaling of shot-noise limit. Under the same squeezing
strength and displacement, it is not strange that the accuracy limit of
parameter estimation is overall inferior that the shot-noise limit when $%
\phi -\theta /2=k\pi \left( k\in 
\mathbb{Z}
\right) $, notably, such conclusions presented both in Fig. \ref{fig:3}(a) and \ref{fig:3}(b)
are consistent with our intuitive prediction.

Meanwhile, we also explore the larger squeezing strength $r=1.5$ under the
same displacement in Fig. \ref{fig:3}(c). It is clearly found that the accuracy limit of parameter estimation in the situation of $\phi -\theta /2=k\pi /2\left( k\in 
\mathbb{Z}
\right) $ will not only break the shot-noise limit but also approach the
Hofmann scaling, even when the phase sensitivity condition is not favorable (%
$\phi -\theta /2=k\pi \left( k\in 
\mathbb{Z}
\right) $), the accuracy limit can likewise beat the
shot-noise limit, which may somewhat counterintuitively. The possibility
arises because QFI describes how rapidly quantum fidelity changes between
two infinitesimally different states. When the squeezing strength becomes
larger, although the error ellipse radius of the DSV state surpasses
coherent state under a bad phase-sensitive condition, QFI is more sensitive
to the parameter, infinitesimal changes may lead the QFI rapidly changed
after parametrical encoding, the combination of which results in the
estimated accuracy limit has superior sensitivity compared with the shot-noise
limit. The conclusions constituted through adjusting $r=1.5$, $\left\vert \alpha\right\vert =10$ in Fig. \ref{fig:3}(d) are the same as what we
achieve in Fig. \ref{fig:3}(b).

Moreover, with the squeezing strength increased to $r=2.5$ (illustrate in
Fig. \ref{fig:3}(e) and \ref{fig:3}(f)), a smaller displacement may enable the DSV state to lose
its phase-sensitive advantages. At this moment, the accuracy limit of
parameter estimation has a better approach to the Hofmann limit regardless of its
phase-sensitive characteristics. Comparing the magnitudes of the different
displacement $\left\vert \alpha \right\vert $ of the DSV states, the state
with the largest $\left\vert \alpha \right\vert $ performs the best on the
estimation because of its the largest error contour angle $\Delta \vartheta $.

\begin{figure*}[h]
\centering
\includegraphics[width=0.75\linewidth]{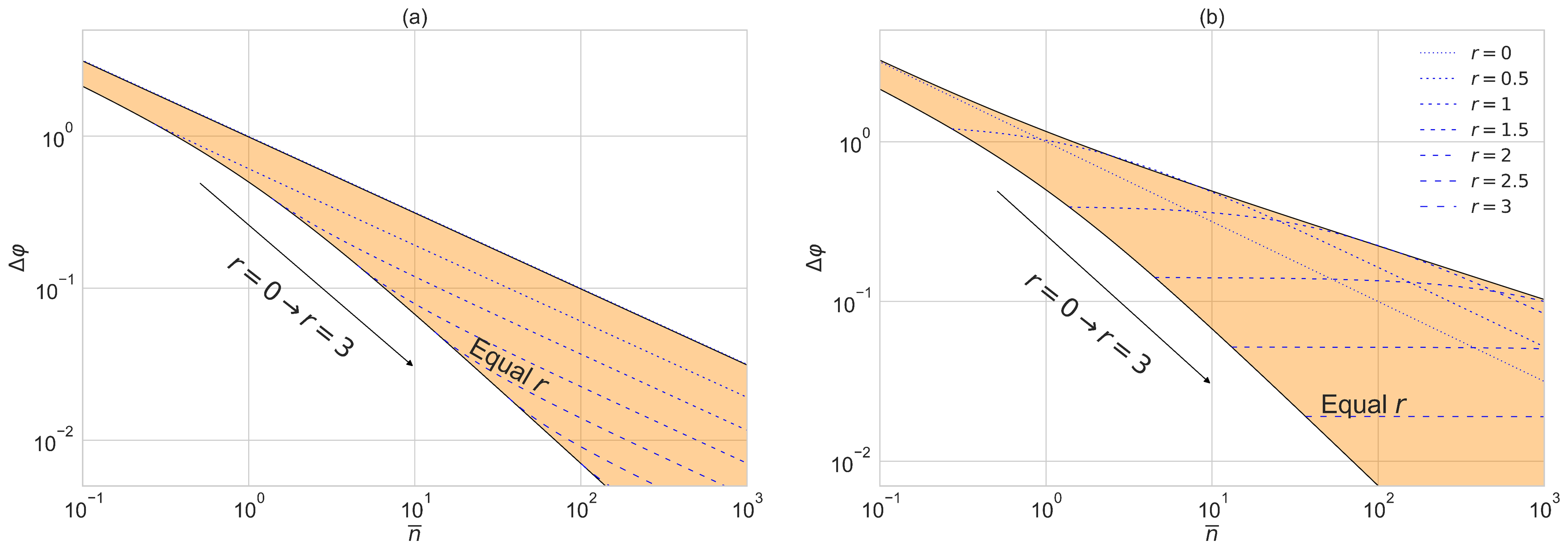}
\caption{Accuracy limit of parameter estimation as a function of the average photon number $\overline{n}$ for any $\left\vert \alpha \right\vert $ and $r$ under different typical phase-sensitive conditions (a) $\phi -\theta /2=\pi /2$\ (b) $\phi -\theta /2=0$. The shaded region represents all possible estimation error for any combination of $\left\vert \alpha \right\vert $ and $r$. The different spacing dashed lines represent the DSV state with the equal $r$ but $\overline{n}$ varying.}
\label{fig:4}
\end{figure*}

\subsection{Under the same average photon number}

\begin{figure*}[htbp]
\centering
\includegraphics[width=\linewidth]{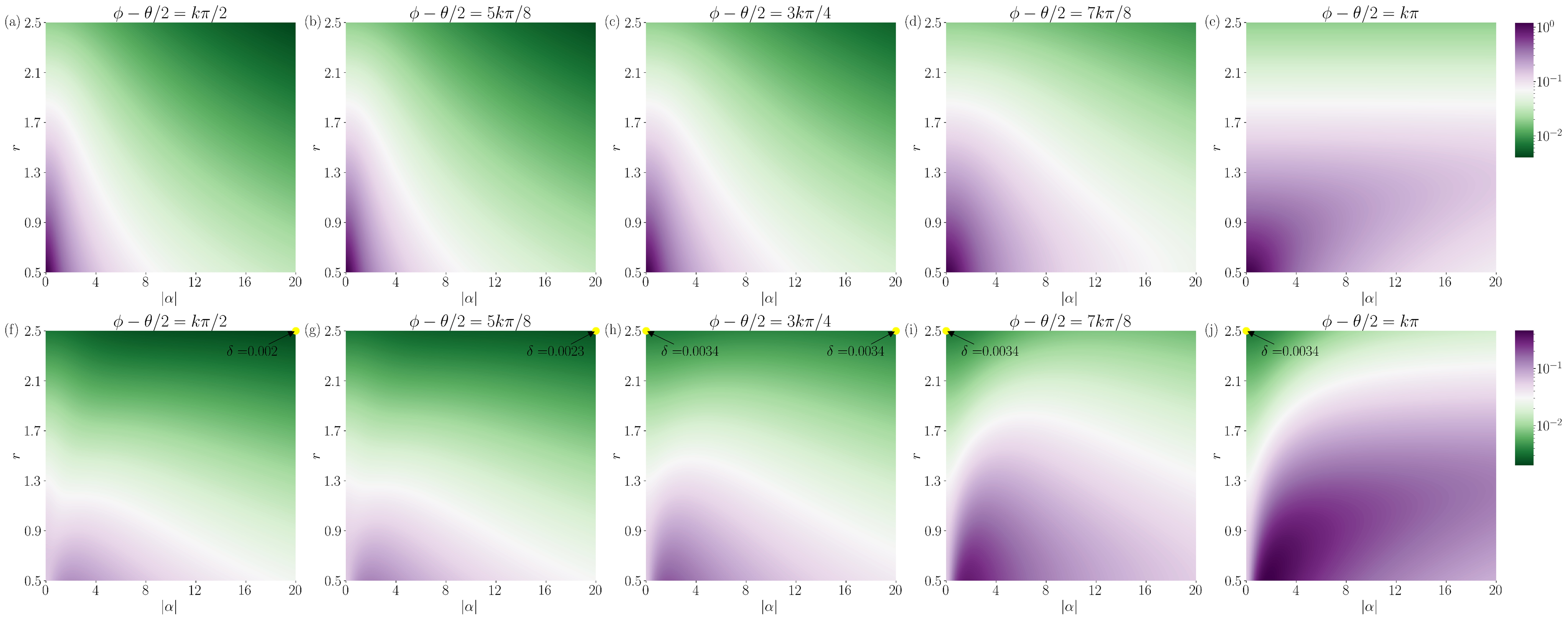}
\caption{The density plots represent the accuracy limit of parameter estimation ((a)--(e)) and the difference between the accuracy limit and the Hofmann limit $(\delta \equiv \Delta \varphi -\Delta \varphi _{Hof})$ ((f)--(j)) as a function of $\left\vert \alpha \right\vert $ and $r$ under different phase-sensitive conditions (a), (f) $\phi -\theta /2=k\pi /2 \left( k\in \mathbb{Z} \right)$; (b), (g) $\phi -\theta /2=5k\pi /8 \left( k\in \mathbb{Z} \right)$; (c), (h) $\phi -\theta /2=3k\pi /4 \left( k\in \mathbb{Z}  \right)$; (d), (i) $\phi -\theta /2=7k\pi /8 \left( k\in \mathbb{Z} \right)$; (e), (j) $\phi -\theta /2=k\pi \left( k\in \mathbb{Z} \right)$. Optimal points which have
the most approach to the Hofmann limit have been labeled in each plot by solid
yellow dots. The colorbar for all density plots in each row are the same.}
\label{fig:5}
\end{figure*}

We illustrate in Fig. \ref{fig:4} how the Cramer-Rao lower bound changes with respect
to the average photon number $\overline{n}$ for any possible squeezing
strength and displacement. The different spacing dashed lines represent the
DSV state with the equal $r$ but $\overline{n}$ varying, which indicates
that increasing the DSV's squeezing strength can constantly improve the
accuracy limit of parameter estimation. The upper boundary of the shaded
area in Fig. \ref{fig:4}(a) stands for the shot-noise limit, corresponding to the
shortest spacing dashed line in Fig. \ref{fig:4}(b). In the $\phi -\theta /2=\pi /2$
regime, (Fig. \ref{fig:4}(a), representing the phase-sensitive condition of any
integer multiple of $\pi /2$), the DSV state with the larger squeezing
strength achieves the higher parameter estimation accuracy as $\overline{n}$
increases, which coincides with our qualitatively geometric intuitive
prediction.

Moreover, as compared with the results acquired in the favorable
phase-sensitive condition, we find in Fig. \ref{fig:4}(b) (i.e., in the $\phi -\theta
/2=0$ regime, representing the phase-sensitive condition of any integer
multiple of $\pi $) that the accuracy limit of parameter estimation is the same as
the circumstance presented in Fig. \ref{fig:4}(a) when the DSV state with no
displacement. However, this superiority of the squeezed-vacuum state
gradually decreases as $\left\vert \alpha \right\vert $ increases. Notably,
we observe that for the DSV state with different squeezing strengths, the
Cramer-Rao lower bound hardly varies with $\overline{n}$ and still beats the
shot-noise limit within a specific photon number range (e.g., when $r=1.5$,
the DSV state with $\overline{n}\in \left( 0,55\right] $ yields superior
performance than the shot-noise limit, which is not shown in Fig. \ref{fig:4}.). When $%
\left\vert \alpha \right\vert $ increases until the accuracy limit of
parameter estimation exceeds the shot-noise limit, parameter estimation
employing the DSV state under this phase-sensitive condition may obtain much
less effective than that of the coherent state.

\subsection{Under the same squeezing strength and displacement}

In order to give a general analysis about how displacement and squeezing
parameter interplay the accuracy limit of parameter estimation and find what
a pair of displacement and squeezing parameter gives the optimal
parameter estimation accuracy, we present the density plot of $\Delta \varphi$ as a function of $\left\vert \alpha \right\vert $ and $r$ in the top row of Fig. \ref{fig:5},
manifesting the Cramer-Rao limit monotonically decreases as $\left\vert
\alpha \right\vert $ and $r$ increase. Besides, we employ the Hofmann limit
as a benchmark to measure the optimality of intrinsic DSV's parameters. The
bottom row of Fig. \ref{fig:5} presents the difference density plot between the
Cramer-Rao limit and the Hofmann limit under some typical phase-sensitive
conditions.

Based on the comparison of five plots in the top row of Fig. \ref{fig:5}, we conclude
the results from the parameter estimation error evolution: the DSV state
with the larger squeezing strength and displacement may achieve a lower
estimation error of parameter except Fig. \ref{fig:5}(e) (i.e., in the $\phi -\theta
/2=k\pi \left( k\in 
\mathbb{Z}
\right) $ regime). We observe in Fig. \ref{fig:5}(e) that when $r$ beyond a certain
range, the increase of $\left\vert \alpha \right\vert $ might not enhance
the parameter estimation performance (e.g., $r=2$). Besides, we also find
that when $\left\vert \alpha \right\vert $ is fixed and surpasses the
specific threshold, the DSV state with the larger $r$ might not obtain a
higher estimation accuracy (e.g., $\left\vert \alpha \right\vert =10$).
These possibilities can be geometrically explained by the compromise of the
smaller error ellipse radius and larger fidelity susceptibility\cite{c47,c48}
discussed in Fig. \ref{fig:3}(c).

The optimal measurement accuracy of parameter estimation is marked at each
density plot in the bottom row of Fig. 5 with solid yellow dots. As depicted
in Fig. 5(h) to (j), we indicate that the DSV's optimal estimation accuracy is
obtained at the left upper corner, where $\left\vert \alpha \right\vert =0$.
Especially, the right upper corner of Fig. 5(h) is also the optimal point
that most approaching the Hofmann limit, which coincides with the
conclusions discussed in Ref. \cite{c49}. Nevertheless, we observe in Fig.
5(f) and (g) that the optimality of intrinsic DSV's parameters will be
transfer to the right upper corner of the density plot when $\phi -\theta
/2\in \left[ k\pi /2,3k\pi /4\right) \left( k\in 
\mathbb{Z}
\right) $, which means that the DSV state with a better phase-sensitive
characteristic and larger squeezing strength and displacement can
significantly enhance the parameter estimation accuracy to the ultimate
quantum limit.

\section{CONCLUSIONS}
\label{sec:4}
In this paper, we first qualitatively assess the performance of parameter
estimation under different phase-sensitive conditions from the perspective
of a general quantum state's geometric error ellipse. Secondly, we explore
the phase-sensitive nonclassical properties in quantum metrology using a
single-mode DSV state. Similar to our geometric intuitive prediction, we
found that the accuracy limit of DSV's parameter estimation is a function of
the phase-sensitive parameter $\phi -\theta /2$ with period $\pi $. Besides,
we also demonstrate that in the $\phi -\theta /2=k\pi /2\left( k\in 
\mathbb{Z}
\right) $ regime, the DSV state with the larger displacement $\left\vert
\alpha \right\vert $ and squeezing strength $r$ can significantly reduce the
estimation error, whereas when $\phi -\theta /2=k\pi \left( k\in 
\mathbb{Z}
\right) $, the DSV state with equal $\left\vert \alpha \right\vert $ and
larger $r$ or equal $r$ and larger $\left\vert \alpha \right\vert $ do not
necessarily enhance the parameter estimation performance. We then
demonstrate that under this circumstance, the selection of the smaller $%
\left\vert \alpha \right\vert $ combined with larger $r$ can beat the
shot-noise limit within a certain average photon number range.

Our initial aim is to reveal how the accuracy limit of parameter estimation
changes with the DSV's phase-sensitive parameter and find what pair of DSV's
intrinsic parameters gives the optimal parameter estimation accuracy under
some typical phase-sensitive conditions. We indicate that the DSV state with
the larger $r$ and $\left\vert \alpha \right\vert $ can obtain the parameter
estimation accuracy that is closest to the ultimate quantum limit when $\phi
-\theta /2\in \left[ k\pi /2,3k\pi /4\right) \left( k\in 
\mathbb{Z}
\right) $, whereas $\phi -\theta /2\in \left( 3k\pi /4,k\pi \right] \left(
k\in 
\mathbb{Z}
\right) $, the same effect can be obtained only when the DSV state
degenerates to a squeezed-vacuum state.\\

\noindent\textbf{Funding.} Anhui Provincial Natural Science Foundation (Grant No. 1908085QA37); National Natural Science Foundation of China (Grant No. 11904369); State Key Laboratory of Pulsed Power Laser Technology Supported by Open Research Fund of State Key Laboratory of Pulsed Power Laser Technology (Grant No. 2019ZR07).\\

\noindent\textbf{Acknowledgment.} We thank the anonymous referee whose comments significantly improved the presentation of this paper.\\

\noindent\textbf{Disclosures.} The authors declare no conflicts of interest.\\


\ifthenelse{\equal{\journalref}{aop}}{%
\section*{Author Biographies}
\begingroup
\setlength\intextsep{0pt}
\begin{minipage}[t][6.3cm][t]{1.0\textwidth} 
  \begin{wrapfigure}{L}{0.25\textwidth}
    \includegraphics[width=0.25\textwidth]{john_smith.eps}
  \end{wrapfigure}
  \noindent
  {\bfseries John Smith} received his BSc (Mathematics) in 2000 from The University of Maryland. His research interests include lasers and optics.
\end{minipage}
\begin{minipage}{1.0\textwidth}
  \begin{wrapfigure}{L}{0.25\textwidth}
    \includegraphics[width=0.25\textwidth]{alice_smith.eps}
  \end{wrapfigure}
  \noindent
  {\bfseries Alice Smith} also received her BSc (Mathematics) in 2000 from The University of Maryland. Her research interests also include lasers and optics.
\end{minipage}
\endgroup
}{}

\end{document}